\documentclass[%
 aip,
 amsmath,amssymb,
 reprint,%
]{revtex4-2}

\usepackage{graphicx}
\usepackage{dcolumn}
\usepackage{bm}
\usepackage{color}
\usepackage[T1]{fontenc}
\usepackage{booktabs}
\usepackage[margin=3cm]{geometry}
\usepackage{algorithm2e}\RestyleAlgo{ruled}
\begin{document}

\title{Predicting chaotic statistics with unstable invariant tori}

\author{Jeremy P. Parker}
\email[Author to whom correspondence should be addressed: ]{jeremy.parker@epfl.ch}
\author{Omid Ashtari}
\author{Tobias M. Schneider}
\affiliation{
Emergent Complexity in Physical Systems Laboratory (ECPS), \'Ecole Polytechnique F\'ed\'erale de Lausanne, CH-1015 Lausanne, Switzerland
}

\begin{abstract}
It has recently been speculated that statistical properties of chaos may be captured by weighted sums over unstable invariant tori embedded in the chaotic attractor of hyperchaotic dissipative systems; analogous to sums over periodic orbits formalized within periodic orbit theory. 
Using a novel numerical method for converging unstable invariant 2-tori in a chaotic PDE, we identify many quasiperiodic, unstable, invariant 2-torus solutions of a modified Kuramoto-Sivashinsky equation exhibiting hyperchaotic dynamics with two positive Lyapunov exponents. The set of tori covers significant parts of the chaotic attractor and weighted averages of the properties of the tori -- with weights computed based on their respective stability eigenvalues -- approximate statistics for the chaotic dynamics. 
These results are a step towards including higher-dimensional invariant sets in a generalized periodic orbit theory for hyperchaotic systems including spatio-temporally chaotic PDEs.

\end{abstract}

\maketitle

\begin{quotation}
Periodic orbit theory formalizes the idea that, if one can identify a large number of non-chaotic unstable periodic orbits embedded within a chaotic attractor, the properties of each of these non-chaotic time-invariant solutions can be summed with suitable weights to predict statistical properties of the chaotic dynamics itself. However, it has been conjectured that in so-called `hyperchaotic' systems, such as turbulent fluid flows and other spatio-temporally chaotic problems, it may be advantageous to consider the higher-dimensional invariant structures called invariant tori, instead of unstable periodic orbits.
Considering a particular hyperchaotic partial differential equation, we here show that one can indeed successfully identify many unstable invariant tori and describe chaotic statistics as sums over the tori.
\end{quotation}

\section{Introduction}

Chaotic dynamics arise naturally from simple interactions in many physical systems, from fluid dynamics to electrical circuits and nonlinear optics. Studying the chaotic dynamics in terms of unstable non-chaotic \emph{invariant} solutions to the underlying evolution equations, which are embedded within the stable chaotic attractor, provides key insights into the observed physics. In the absence of special symmetries, two types of unstable invariant solutions are generally studied: equilibria, zero-dimensional unstable fixed points in the state space of the system; and periodic orbits, non-chaotic time-periodic solutions corresponding to one-dimensional closed loops in state space. 
Though equilibria are usually rare, in a general continuous dynamical system, periodic orbits are expected to be dense within the attractor \citep{pugh1967improved}.
Trajectories within chaotic attractors closely shadow unstable periodic orbits, and consequently, periodic orbits are often described as the `backbone' of chaos. 
For systems with dense periodic orbits, the construction of dynamical zeta functions \citep{bogomolny1992dynamical} allows statistical quantities to be evaluated as sums over the collection of all periodic orbits  \citep{cvitanovic1988invariant, eckhardt1994periodic, chandler2013invariant}.

However, for dissipative systems the `closing lemmas' \citep{anosov2012closing} which imply density of periodic orbits have not been extended to include the very smooth dynamical systems which arise from physical laws. Indeed, the formal lemmas certainly do not apply in the non-hyperbolic case where a system has multiple zero Lyapunov exponents in addition to a positive one, for example because of a continuous symmetry. In this case, instead of periodic orbits we find \emph{relative} periodic orbits \citep{budanur2017relative}, which are in fact a special case of invariant 2-tori.

Invariant 2-tori are, after equilibria and periodic orbits, the next simplest invariant solutions in continuous dynamical systems. They represent quasiperiodic behaviour, in which two different fundamental frequencies interact.
In a previous paper \citep{parker2022invariant}, it was shown that for a particular dissipative system of ordinary differential equations which exhibits hyperchaos with three positive Lyapunov exponents, invariant 2-tori are generically found embedded within the attractor. This led to the conjecture that unstable invariant tori (UITs), as well as unstable periodic orbits (UPOs), should be taken into account to describe dissipative hyperchaotic systems, and that UITs could be used to predict statistics of interest in such systems.

One important class of hyperchaotic dissipative systems are the spatio-temporally chaotic partial differential equations (PDEs) which arise in fluid dynamics and related fields.
Generic invariant tori could allow the study of key phenomenology within these systems, for which periodic orbits are rare or at least computationally problematic to detect. For example, in wall-bounded turbulence, it has proven difficult to find periodic orbit solutions which capture the interaction between different processes at different length-scales \citep{doohan2022state}. In the absence of phase locking, we expect the different temporal frequencies associated with these length-scales to lead to the dynamics manifesting as invariant tori rather than periodic orbits.
It is conjectured that as the complexity of spatio-temporal chaos increases, which is associated with a greater number of positive Lyapunov exponents, the likelihood of finding and relative importance of UITs increases. However, a method to find such tori generically remains elusive.

This paper aims at advancing the methodology for generically finding and exploiting UITs in the description of chaos. We consider the PDE system of a forced generalised Kuramoto-Sivashinsky equation (gKSE). 
Here, we can find invariant tori through their connection to relative periodic orbits in the unforced case. We show that UITs are plentiful and can be summed over to predict various average quantities of the chaotic dynamics.
The paper is laid out as follows: in section \ref{sec:equations} we present both the unforced and forced gKSE, and the Lyapunov exponents associated with our chosen parameters; in section \ref{sec:tori} we present and apply methods to find UITs, which are used in section \ref{sec:stats} to predict statistics for the system; and the results are discussed in section \ref{sec:conclusion}.

\section{The forced generalised KSE}
\label{sec:equations}
\begin{figure*}
    \centering
    \includegraphics[width=\textwidth]{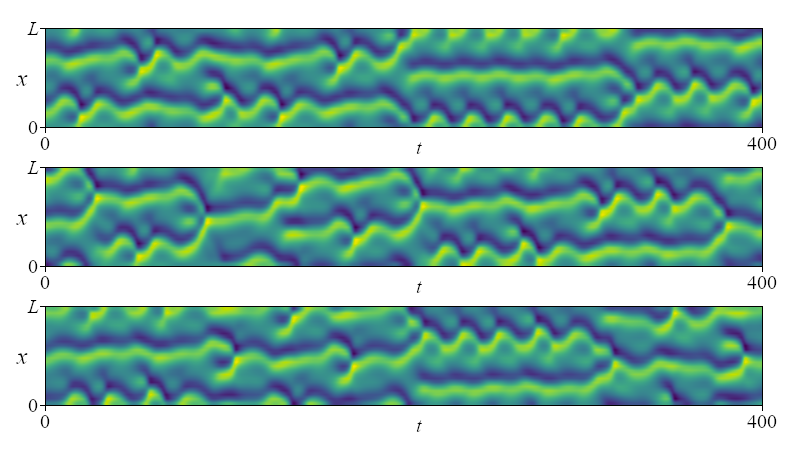}
    \caption{Typical chaotic timeseries at $L=22$ of length $t=400$. Top: the classic KSE $\beta=0$, $\epsilon=0$. Middle: The gKSE with $\beta=0.01$ but $\epsilon=0$. Bottom: Our forced gKSE with $\beta=0.01$ and $\epsilon=0.002$. The behaviour appears very similar, despite the fact that the first system has two symmetries which the last does not.}
    \label{fig:timeseries}
\end{figure*}

The one-dimensional generalised Kuramoto-Sivashinsky equation (gKSE)  \citep{kudryashov1990exact,khater2008numerical,lai2009lattice,lakestani2012numerical} for a real-valued scalar field $u(x,t)$ defined on a periodic spatial domain of size $L$ can be written as
\begin{equation}
\label{eq:gKSE}
    \partial_t u + u \partial_x u + \partial_x^2 u + \beta \partial_x^3 u + \partial_x^4 u = 0.
\end{equation}
For $\beta=0$, the gKSE reduces to the classic Kuramoto-Sivashinsky equation (KSE), which has many applications in physics \citep{laquey1975nonlinear,sivashinsky1977nonlinear,chang1994wave}.
A non-zero value of $\beta$ acts to break the discrete left-right antisymmetry of the KSE, but for $\beta=0.01$, the observed dynamics is not altered significantly. The significance of breaking the discrete symmetry will be discussed in section \ref{sec:LEs}.
The complexity of the dynamics in general increases as the domain size $L$ is increased. We here consider $L=22$, for which the dynamics are chaotic.
Typical timeseries of the classic KSE and the gKSE for the considered values of the control parameters are shown in figure \ref{fig:timeseries}.
The mean flow $\int_0^L u\, \mathrm{d}x$ is a conserved quantity of (\ref{eq:gKSE}), and without loss of generality can be taken to be zero.

Equation (\ref{eq:gKSE}) is invariant under the continuous family of transformations $u(x+l,t)\mapsto u(x,t)$, parameterised by $l\in\mathbb{R}$. 
This has the consequence that periodic orbits are unlikely in the chaotic attractor, and instead \emph{relative} periodic orbits (RPOs), which have a non-zero phase velocity, are readily found.
To break the continuous symmetry of (\ref{eq:gKSE}) in a controlled way, we add a forcing term and instead considered the forced gKSE,
\begin{equation}
\label{eq:fgKSE}
    \partial_t u + u \partial_x u + \partial_x^2 u + \beta \partial_x^3 u + \partial_x^4 u = \epsilon \sin {\left(2\pi x/L\right)},
\end{equation}
where $\epsilon$ is a small parameter, which controls the breaking of the continuous shift symmetry. Beyond the control of the continuous symmetry, there is no physical motivation for studying this augmented system; the choice will become clear below. We will use $\epsilon$ on the order of $10^{-3}$, which is sufficiently large that the system is detectably different from the unforced equation, without being so large that the qualitative dynamics significantly change. See figure \ref{fig:timeseries} for a visual comparison of the dynamics with and without forcing.

\begin{figure*}
    \centering
    \includegraphics[width=\textwidth]{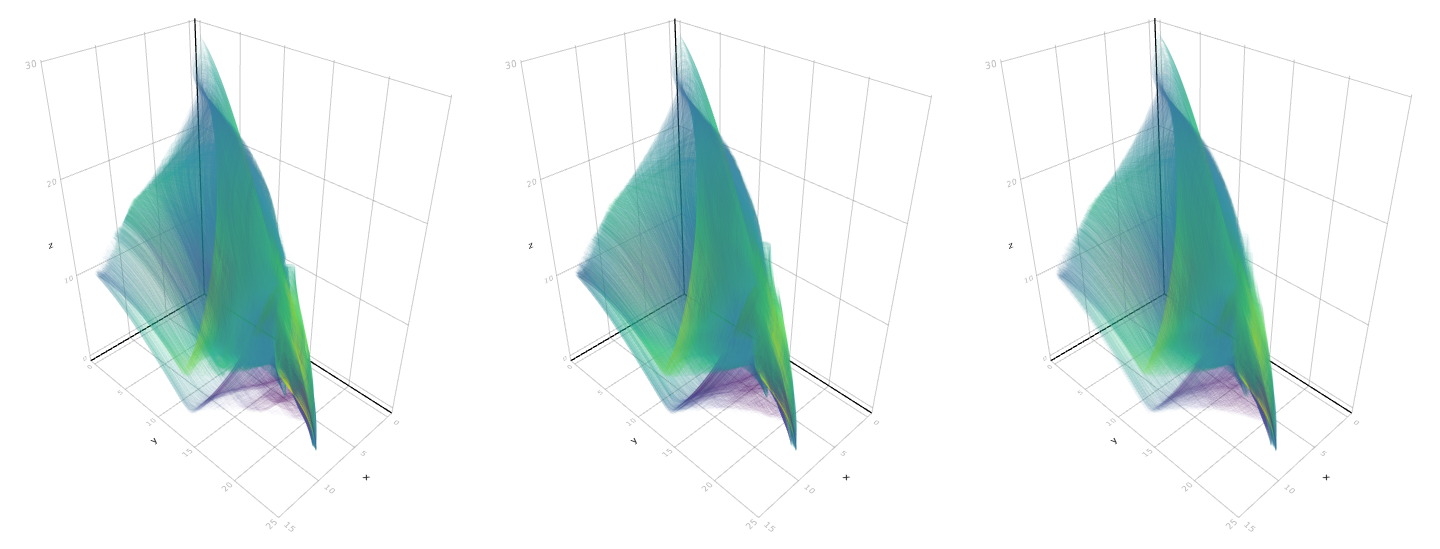}
    \caption{The chaotic attractor at $L=22$, depicted as a timeseries of length $t=10^6$. Left: the classic KSE $\beta=0$, $\epsilon=0$. Middle: The gKSE with $\beta=0.01$ but $\epsilon=0$. Right: Our forced gKSE with $\beta=0.01$ and $\epsilon=0.002$. 
    The projection shows, in each case, the magnitude of the first, second and third Fourier coefficients on the $x$, $y$ and $z$ axes respectively. Notice the subtle effect of nonzero $\beta$ and $\epsilon$ to `blur' the attractor.}
    \label{fig:chaoticattractors}
\end{figure*}

Figure \ref{fig:chaoticattractors} shows a projection of the chaotic attractors at the discussed parameter values yielding the classical KSE, the unforced gKSE and the forced gKSE. The choice of projection is important here -- since $\epsilon\neq0$ breaks the continuous phase-shift symmetry we choose a projection which factors out this phase shift, namely the absolute value of the complex Fourier coefficients.

We time-march these PDEs using an exponential time-differencing fourth-order Runge-Kutta scheme, following \citet{kassam2005fourth}. 
This is coded in Julia so that we may find Jacobians through automatic differentiation using the package Zygote \citep{innes2018don}. Some additional calculations were performed using an identical algorithm in MATLAB. Throughout, we discretise the domain with $24$ points, or $N=16$ Fourier modes after 2/3 dealiasing, which is sufficient for this relatively low value of $L$.

From (\ref{eq:fgKSE}) we define the instantaneous energy production
\begin{equation}
    \label{eq:production}
    P := \int_0^L \left[ \left(\partial_x u\right) ^2  +\epsilon u \sin{\left( 2\pi x/L\right)} \right]\mathrm{d}x,
\end{equation}
and dissipation
\begin{equation}
    \label{eq:dissipation}
    D := \int_0^L \left(\partial^2_x u\right) ^2\mathrm{d}x,
\end{equation}
which we will use as key statistics of the flow. The long-time averages of these should be equal for a statistically stationary trajectory, and they will be equal when averaged over any invariant solution.

\subsection{Lyapunov exponents}
\label{sec:LEs}
The average rate of growth or decay of perturbations to a trajectory within a chaotic attractor is measured by the Lyapunov exponents. The presence of at least one positive Lyapunov exponent is a necessary and sufficient condition for an attractor to be chaotic. A chaotic attractor is called hyperchaotic if its Lyapunov spectrum contains at least two positive exponents.

We compute the leading Lyapunov exponents using algorithm \ref{algo:Lyapunov_exponents}, which is inspired by an approach initially proposed by \citet{Benettin1980}. In this algorithm, the state $u$ and the orthonormal basis $\mathbf{Q}$ are mapped to $f^\tau(u)$ and $\mathbb{J}_{u}^\tau\mathbf{Q}$, respectively, where $f^\tau$ is the time evolution operator for time $\tau$ and $\mathbb{J}_{u}^\tau = \nabla_{u}f^\tau(u)$ is the Jacobian of the flow. The deformed basis is reorthonormalized via QR decomposition, which at the same time allows us to extract the finite-time evaluation of the Lyapunov exponents $\chi_i$ from the diagonal elements of the right triangular matrix\cite{Eckmann1985}. This process is repeated for $n\gg1$ times, and the Lyapunov exponents are averaged. Prior to this loop, the dynamics is integrated for a sufficiently long time $\tau_0$ to ensure that the trajectory is confined to the chaotic attractor.

\begin{algorithm}
\label{algo:Lyapunov_exponents}
\caption{Computation of the leading Lyapunov exponents.}
\SetKwInOut{Input}{Input}
\SetKwInOut{Output}{Output}
\Input{Forced gKSE parameters $L$, $\beta$ and $\epsilon$\\
Initial condition IC ($N$-dim. state) \\
Transition time $\tau_0$\\
Number of Lyapunov exponents $p$\\
Reorthonormalization time $\tau$\\
Number of reorthonormalizations $n$}
\Output{$p$ leading Lyapunov exponents $\mathbf{X}=[\chi_1\;\dots\;\chi_p]^\top$}
\vspace{2mm}\hrule\vspace{1mm}
$u\gets f^{\tau_0}(\mathrm{IC})$\tcp*{time marching from IC}
$\tilde{\mathbf{Q}} \gets \left[\mathbf{I}_p \;|\; \mathbf{0}_{p\times(N-p)}\right]^\top$\;
$\tilde{\mathbf{Q}}_{:,1}\gets\partial_t u$\;
\If{$\epsilon=0$} {
    $\tilde{\mathbf{Q}}_{:,2}\gets\partial_x u$\;
}
QR decompose $\tilde{\mathbf{Q}}=\mathbf{Q}\mathbf{R}$\;
$\mathbf{X}\gets\mathbf{0}_{p\times1}$\;
\For{$j=1$ to $n$}{
    $\mathbf{W} \gets \mathbb{J}_{u}^\tau\mathbf{Q}$ \;
    $u\gets f^\tau(u)$\;
    $\mathbf{W}_{:,1}\gets\partial_t u$\;
    \If{$\epsilon=0$} {
    $\mathbf{W}_{:,2}\gets\partial_x u$\;
    }
    QR decompose $\mathbf{W}=\mathbf{Q}\mathbf{R}$\;
    $\mathbf{X}\gets\mathbf{X}+\ln\left(\mathrm{diag}(\mathbf{R})\right)/\tau$\;
}
\Return $\mathbf{X}/n$
\end{algorithm}

\begin{figure}
    \centering
    \includegraphics[width=\columnwidth]{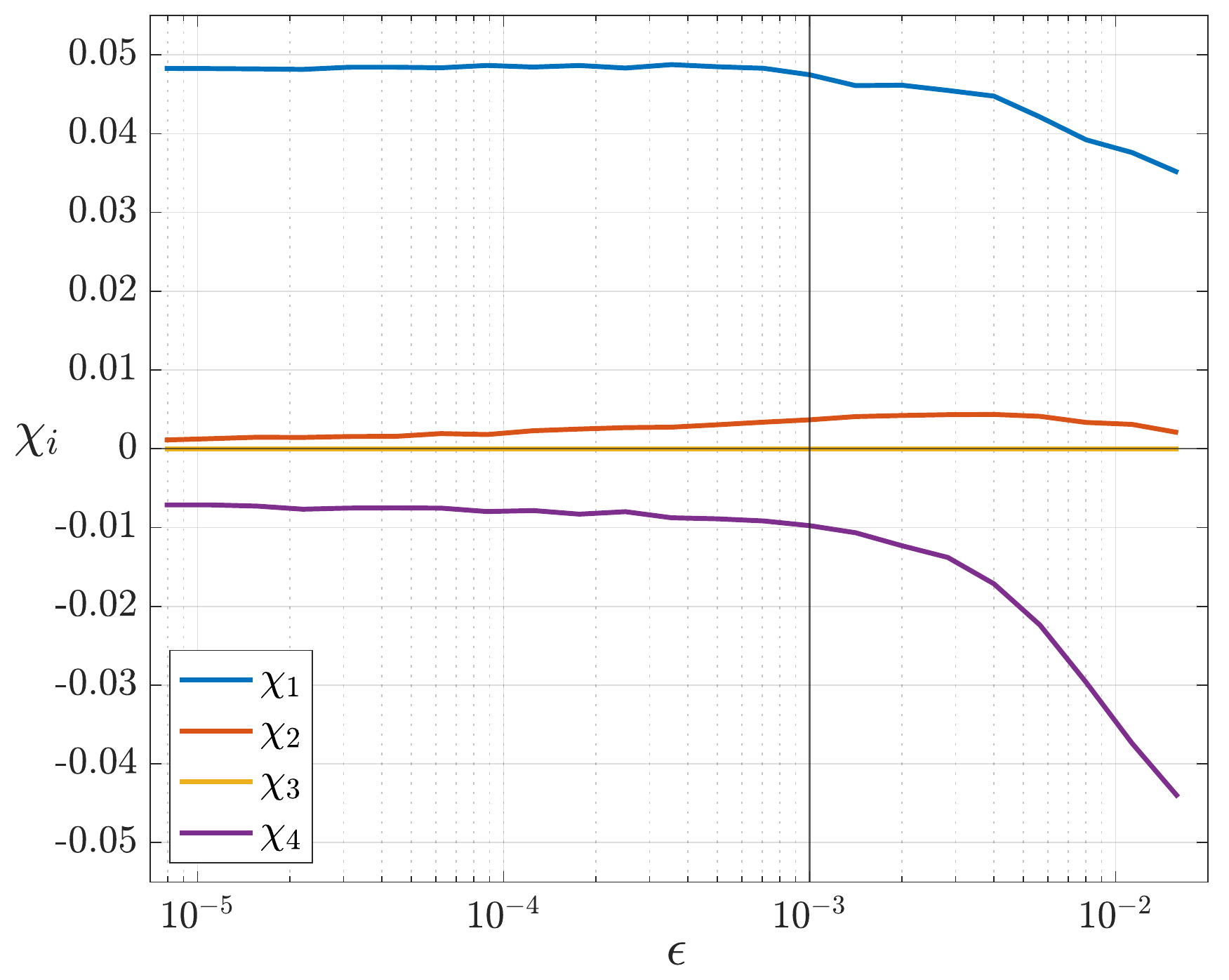}
    \caption{The three most positive Lyapunov exponents of the forced gKSE with $L=22$ and $\beta=0.01$, as $\epsilon$ is varied. The vertical line marks $\epsilon=0.001$.}
    \label{fig:Lyapunov_exponents}
\end{figure}

The four leading Lyapunov exponents of the chaotic attractor over $2^{-7}<10^3\epsilon<2^4$ (with $\beta=0.01$ and $L=22$ being fixed) are shown in Figure \ref{fig:Lyapunov_exponents}. The parameters $\tau_0=2.5\times10^3$, $\tau=2.0$ and $n=7.5\times10^5$ are fixed in the calculations. Outside the plotted range of $\epsilon$, the attractor behaves as follows: for large values of $\epsilon$, the global attractor is a stable fixed point; as $\epsilon$ decreases, the system becomes chaotic via the Ruelle–Takens–Newhouse route to chaos: at $\epsilon\approx0.1202$ the fixed point loses its stability and the attractor becomes a limit cycle; at $\epsilon\approx0.1021$ the attractor becomes a stable torus; and at $\epsilon\approx0.0998$ the attractor turns chaotic and remains chaotic until $\epsilon\approx 0.0819$. Below this value the attractor consists of (possibly coexisting) stable periodic orbits or stable tori. At $\epsilon \approx 0.0190$ the attractor again becomes chaotic and remains so for all smaller values of $\epsilon$.

Due to the presence of an additional zero Lyapunov exponent associated with the continuous translational symmetry in the case of unforced $\epsilon=0$ system, by definition, the chaotic attractor is non-hyperbolic, and we therefore do not expect to find many periodic orbits, but rather relative periodic orbits (RPOs).
For $\beta=0$, the combination of the continuous and discrete symmetries leads to a dense class of periodic orbits called \textit{preperiodic} orbits by \citet{cvitanovic2010state}, but these are not present when $\beta\neq 0$.
As the continuous symmetry is broken by $\epsilon>0$, the remaining RPOs should generically become invariant 2-tori, as discussed in section \ref{sec:continuation}.
Intuitively, for 2-tori to be embedded within a hyperbolic chaotic attractor, we require more than one positive Lyapunov exponent - with only one, the chaotic attractor locally looks like a very thin sheet which cannot geometrically include an invariant 2-dimensional non-chaotic manifold.
Indeed, we see two positive Lyapunov exponents for $\epsilon>0$, so this is consistent.
Note that the Kaplan-Yorke dimension $D_{KY}$ of the attractor slightly \emph{decreases} with increasing $\epsilon$, while the first and fourth Lyapunov exponents become more negative (see Table \ref{tab:Lyapunov_exponents}). In all cases, $D_{KY}\approx4.2$ -- the system we are studying exhibits low-dimensional chaos, in contrast with the spatio-temporal chaos seen at higher $L$ in the KSE, for example.

\begin{table}
    \centering
    \caption{The eight most positive Lyapunov exponents of the chaotic attractor of the forced gKSE for domain length $L=22$ and different parameter settings.}
    \begin{tabular}{p{0.12\linewidth} p{0.2\linewidth} p{0.2\linewidth} p{0.2\linewidth}}
        \toprule
            & $\beta=0$ & $\beta=0.01$ & $\beta=0.01$ \\
            & $\epsilon=0$ & $\epsilon=0$ & $\epsilon=0.001$ \\
        \midrule
            $\chi_1$ & \;0.0484 & \;0.0485 & \;0.0476 \\  
            $\chi_2$ & \;0 & \;0 & \;0.0036 \\
            $\chi_3$ & \;0 & \;0 & \;0 \\
            $\chi_4$ & -0.0028 & -0.0061 & -0.0098 \\
            $\chi_5$ & -0.1884 & -0.1805 & -0.1815 \\
            $\chi_6$ & -0.2562 & -0.2608 & -0.2600 \\
            $\chi_7$ & -0.2902 & -0.2918 & -0.2903 \\
            $\chi_8$ & -0.3102 & -0.3089 & -0.3083 \\
        \midrule
            $D_\mathrm{KY}$ & \;4.2425 & \;4.2348 & \;4.2279\\
        \bottomrule
    \end{tabular}
    \label{tab:Lyapunov_exponents}
\end{table}

\section{Unstable invariant tori}
\label{sec:tori}

\subsection{Torus convergence algorithm}
\label{sec:method}

\begin{figure}
    \centering
    \includegraphics[width=7cm]{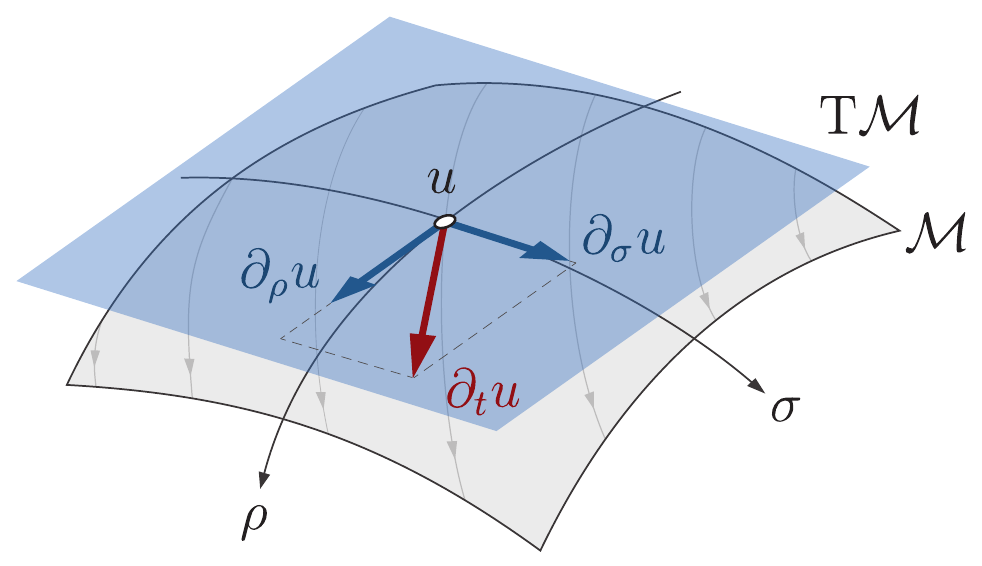}
    \caption{The flow $\partial_t u$ at any point on the surface of the 2-torus $\mathcal{M}$ can be expressed as a linear combination of the tangent vectors along the coordinates $\sigma$ and $\rho$ which span the local tangent space $\mathrm{T}\mathcal{M}$. See equation (\ref{eq:tangent}).}
    \label{fig:tangent_space}
\end{figure}

Rather than discretising a loop on a Poincaré section which slices the torus, as in \citet{parker2022invariant}, the full surface of the 2-torus is parameterised by coordinates $(\rho,\sigma)\in [0,2\pi)\times[0,2\pi)$. The local dynamics on the torus is assumed to be a rotation with a fixed velocity $(R,S)\in \mathbb{R}^2$, so that
\begin{equation}
\label{eq:tangent}
    \partial_t u = R \partial_\rho u + S \partial_\sigma u,
\end{equation}
which states that the flow of the dynamical system lies in the tangent space of the torus at that point, as shown in figure \ref{fig:tangent_space}.
For $u(x,\rho,\sigma)$ to describe an invariant torus of the system (\ref{eq:fgKSE}), we therefore require that
\begin{equation}
\label{eq:torus}
    R \partial_\rho u + S \partial_\sigma u + u \partial_x u + \partial_x^2 u + \beta \partial_x^3 u + \partial_x^4 u = \epsilon \sin {\frac{2\pi x}{L}}.
\end{equation}

The fact that $R$ and $S$ are independent of $\rho$ and $\sigma$ partially constrains the choice of parameterisation, which improves numerical convergence. 
but also assumes quasiperiodic dynamics.
This is a strong assumption which is certainly not valid in Arnold tongues, regions of parameter space in which phase-locking implies the existence of periodic orbits on an invariant torus. In such regions, we may instead directly converge the periodic orbits\citep{parker2022invariant}. In practice, when the phase locking of a UIT is such that the periodic orbits on it are of very long period, it is indistinguishable numerically from a quasiperiodic invariant torus, and our algorithm will converge.

We start with an initial guess which geometrically describes a 2-torus in state space, but not an invariant manifold. We then iteratively deform this torus until (\ref{eq:torus}) is satisfied at every point on the surface.
The torus is discretised with $N=16$ Fourier modes in $x$, as for (\ref{eq:fgKSE}), so the linear terms are entirely local differential operators in Fourier space. In $\rho$ and $\sigma$, we discretise on an evenly-spaced grid of $M\times M= 64\times64$ points, and the derivatives are calculated with dense trigonometric differentiation matrices.
This means that we can compute a sparse Jacobian for the left hand side of (\ref{eq:torus}), of size $NM^2\times(NM^2+2)$, where the two additional columns come from derivatives with respect to $R$ and $S$.
The matrix consists of dense $N\times N$ blocks on the main diagonal, corresponding to the nonlinear terms, and diagonal $N\times N$ blocks elsewhere, corresponding to the derivatives with respect to $\rho$ and $\sigma$. This gives a total of $M^2N^2 + M^4N + M^2N$ nonzero entries.
Following \citet{cvitanovic2010state}, we can solve this system using a Levenberg-Marquardt algorithm. As a consequence, we do not need to introduce additional constraints for this underconstrained optimisation problem. At each iteration of Levenberg-Marquardt, the sparse linear system is solved using the conjugate-gradient-like LSMR \citep{fong2011lsmr}. Since this requires only sparse matrix multiplications, this can be performed on a GPU in Julia at great speed despite the size of the system.

\begin{figure}
    \centering
    \includegraphics[width=0.6\columnwidth]{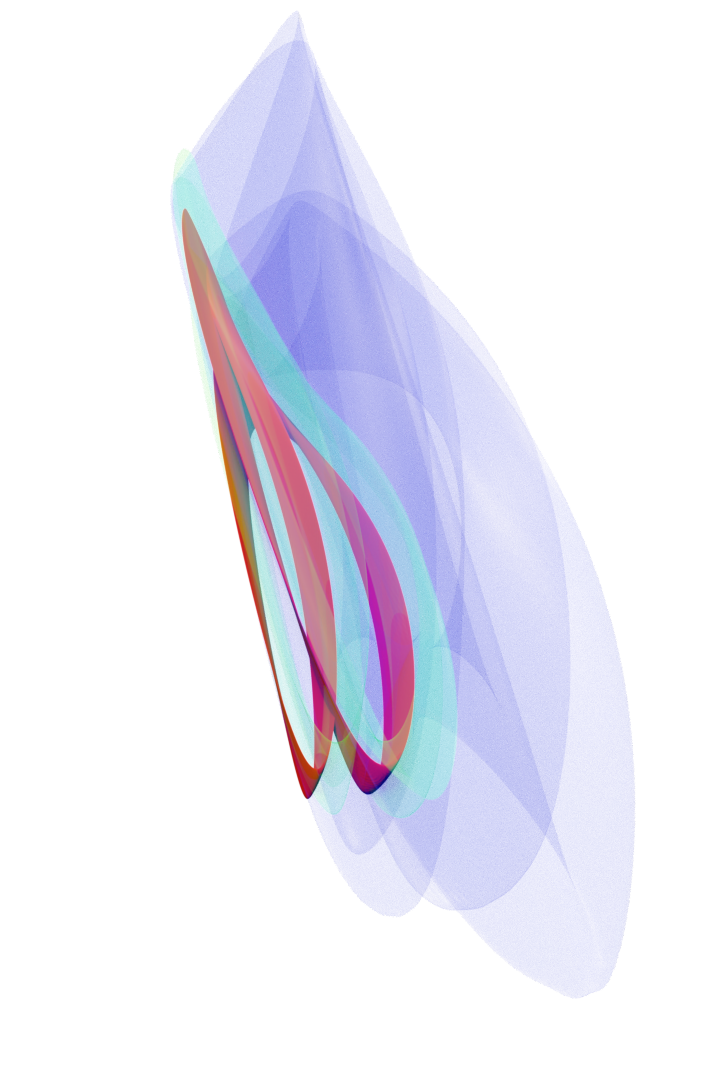}
    \vspace{-0.5cm}
    \caption{Initial guess (blue), the partially converged torus after 5 iterations (green), and the fully converged UIT (red) for $T_2$ at $\epsilon=0.003$. The projection is as per figure \ref{fig:chaoticattractors}.}
    \label{fig:torusconverge}
\end{figure}

As a consequence of using the Levenberg-Marquardt algorithm, our method is able to converge from relatively poor initial guesses, since it is effectively gradient-descent when far from a solution, and so shares the convergence properties of recent improvements in methods for the computation of periodic orbits \citep{azimi2020adjoint,parker2022variational,page2022recurrent}. Figure \ref{fig:torusconverge} shows the torus-shaped but not invariant initial guess at $\epsilon=0.002$ (generated by continuation, see section \ref{sec:continuation}), and the drastically different converged invariant solution.

Our method is also able to converge some UPOs which exist when strong phase-locking is present, as UPOs are a degenerate case of the solution given by (\ref{eq:torus}) in which either $\partial_\rho u \equiv 0$ or $\partial_\sigma u \equiv 0$. Indeed, if they were both zero, the algorithm would have converged to a steady state solution, but this never occurred.

\subsection{Continuation of UITs from RPOs}
\label{sec:continuation}

A relative periodic orbit is a particular case of an invariant torus, where one of the two dimensions of the invariant manifold exactly corresponds to a continuous symmetry of the system. In the presence of such a symmetry, RPOs are expected to be the generic structure in chaotic attractors, with pure periodic orbits rare special cases, for example related to bifurcations of solutions outside the chaos.
The unforced system with $\epsilon=0$ has a continuous symmetry, and we can find RPOs in it using the now-routine method of recurrent flow analysis, followed by a Newton-Krylov based shooting method \citep{chandler2013invariant}. Our implementation of recurrent flow analysis exactly matches that of \citet{cvitanovic2010state}.

Given an RPO with period $T$ and phase velocity $c$ so that $u(x+cT, t+T)=u(x,t)$ for all $x$ and $t$, we can satisfy the form of the previous subsection by defining
\begin{equation}
    u(x,\rho,\sigma) = u\left(x+\frac{L\sigma}{2\pi}, \frac{T\rho}{2\pi}\right), R=\frac{2\pi}{T}, S=-\frac{2\pi c}{L}.
\end{equation}
Therefore, we can take RPOs found via recurrent flow analysis and shooting and reconverge them using the algorithm described in section \ref{sec:method}. Afterwards, we continue the invariant solution to non-zero $\epsilon$, at which point it ceases to be an RPO and is instead a generic 2-torus.

It was found to be sufficient to use the RPO as an initial condition for the algorithm at $\epsilon=0.001$, and then use this converged UIT and the $\epsilon=0$ RPO to linearly extrapolate an initial guess for $\epsilon=0.002$.
Several of the UITs were continued to $\epsilon=0.005$ and beyond, but for the more geometrically complex tori this proved difficult and higher resolution discretisations of $\rho$ and $\sigma$ are likely to be required.
Indeed, we only searched for RPOs for periods up to $T=100$ for this reason.

Figure \ref{fig:torusprojections} shows two different projections of a simple RPO at $\epsilon=0$ which is continued to a UIT at $\epsilon=0.01$. With a projection which `quotients out' the continuous symmetry, the RPO appears as a simple loop but is clearly a full torus at non-zero $\epsilon$. A more naïve projection shows that the RPO is indeed topologically a torus at $\epsilon=0$, but then the effect of $\epsilon$ is obscured.

\begin{figure*}
    \centering
    \includegraphics[width=0.8\textwidth]{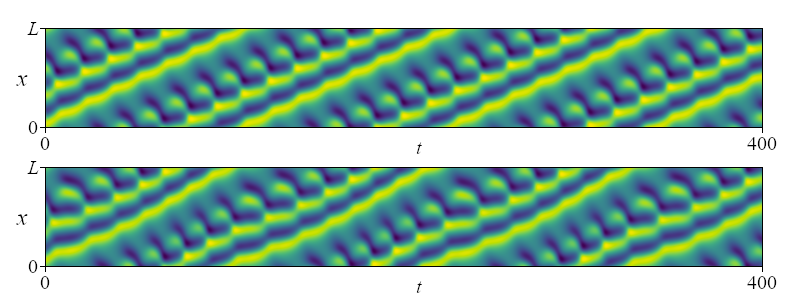}
    \caption{Timeseries at $L=22$, $\beta=0.01$ of length $t=200$ for the solution $T_{10}$. Top: relative periodic orbit at $\epsilon=0$. Bottom: full torus at $\epsilon=0.01$. Note the modulation of the periodic behaviour in the forced, asymmetric case.}
    \label{fig:torustimeseries}
\end{figure*}

\begin{figure*}
\centering
    \includegraphics[width=0.24\textwidth]{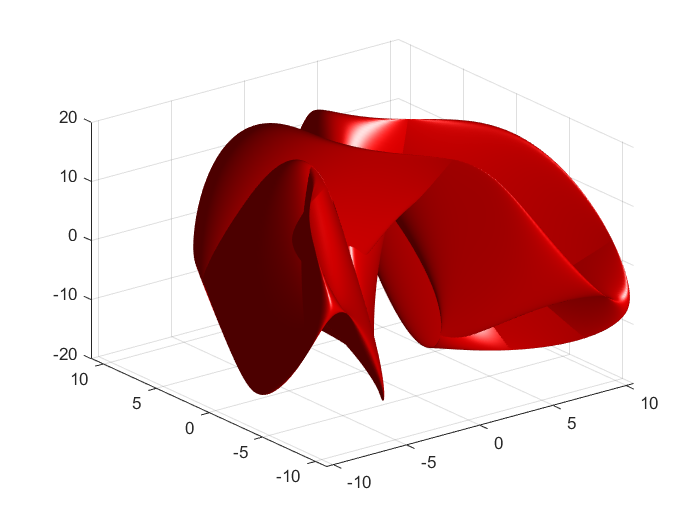}
    \includegraphics[width=0.24\textwidth]{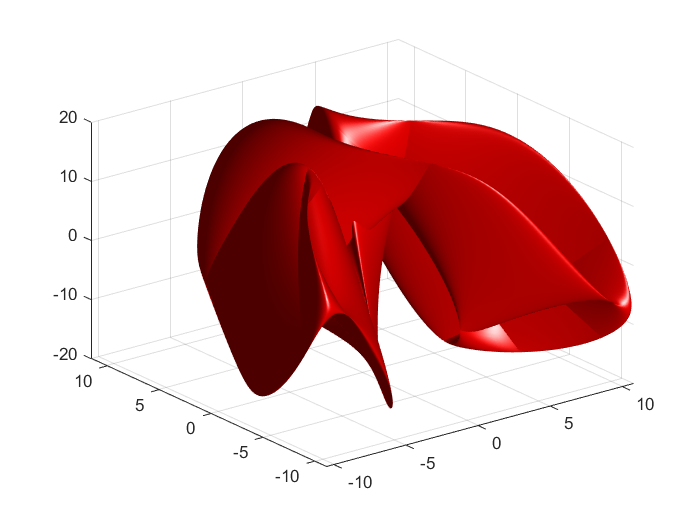}
    \includegraphics[width=0.24\textwidth]{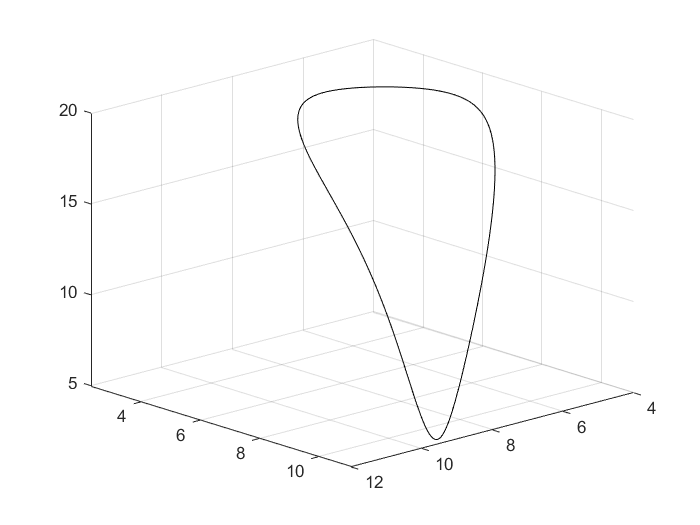}
    \includegraphics[width=0.24\textwidth]{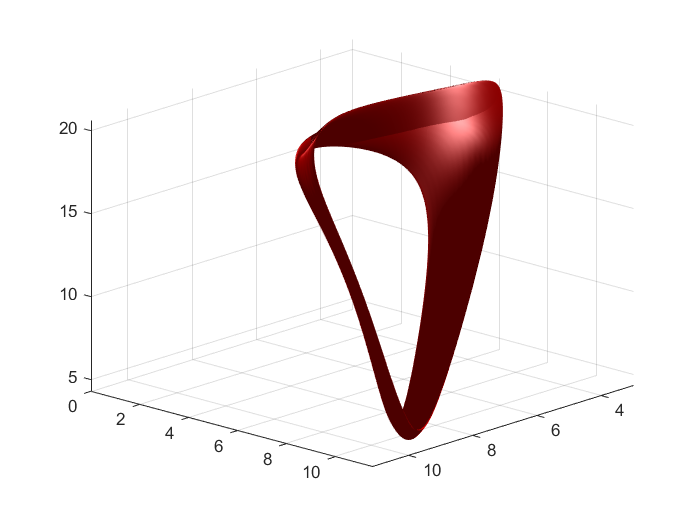}
    \caption{Projections of the solutions of figure \ref{fig:torustimeseries}.
    Left to right: (a) RPO at $\epsilon=0$, showing the \emph{real part} of the first three Fourier coefficients, (b) UIT at $\epsilon=0.01$, real part of first three Fourier coefficients, (c) RPO at $\epsilon=0$, \emph{absolute value} of first three Fourier coefficients, (d) UIT at $\epsilon=0.01$, absolute values.
    The first projection shows that the RPO is indeed geometrically a two-dimensional invariant torus, but disguises the fact that this is due to the continuous symmetry and does not easily allow us to distinguish the two values of $\epsilon$. The second projection, which is that used in the other figures, removes the continuous symmetry and we can clearly see the development from RPO to UIT.}
    \label{fig:torusprojections}
\end{figure*}

\begin{figure}
    \centering
    \includegraphics[width=0.8\columnwidth]{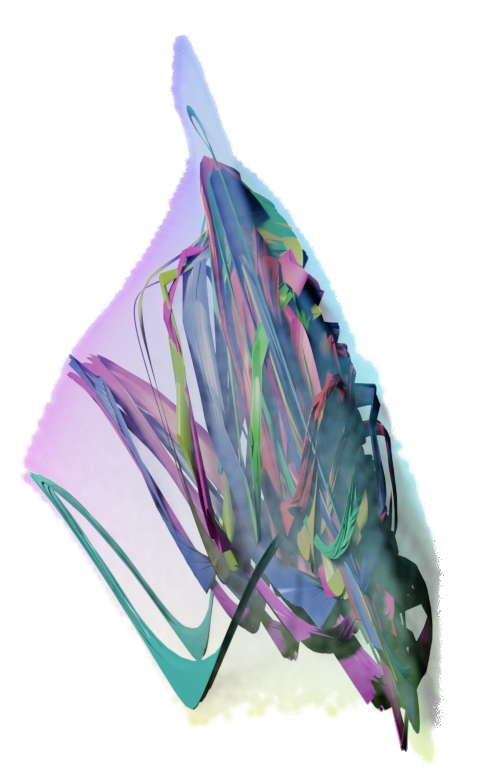}
    \caption{The twelve converged tori at $\beta=0.01$, $\epsilon=0.001$, overlaid on the chaotic attractor (rendered as a cloud). The UITs lie within and capture the fractal structure of the attractor. Projection as per figure \ref{fig:chaoticattractors}.}
    \label{fig:alltori}
\end{figure}

Twelve distinct UITs were successfully converged at $\epsilon=0.001$ from the continuation of 27 distinct RPOs, and eight of these were successfully continued to $\epsilon=0.002$. A projection of all eight of these is shown in figure \ref{fig:alltori}.
In addition to these, a common travelling wave solution at $\epsilon=0$ was continued to give a periodic orbit, but this was omitted from our calculations in the following sections, as were any UPOs resulting from phase-lockings of the UITs.

\subsection{Stability of UITs}
Algorithms for finding the stability properties of invariant tori are notoriously difficult to use \citep{jorba2001numerical, wysham2006iterative}. Here we propose a simple iterative method which accurately finds the leading Lyapunov exponents for a UIT, i.e. the real parts of the stability eigenvalues. It is sufficient to know these to calculate the weights for the statistical averages discussed in section \ref{sec:stats}. Our method does not give the imaginary parts of the eigenvalues, nor the linear manifolds associated with the eigenvectors.

The algorithm is an extension of that discussed in section \ref{sec:LEs} to calculate Lyapunov exponents for the attractor. In theory, if we start with a point exactly on an invariant solution, the Lyapunov exponents calculated using the given algorithm should give us exactly the Lyapunov exponent of the invariant solution. However, as the solutions of interest are all unstable, a trajectory started from a calculated point of the invariant solution will quickly drift away. Consequently, we augment the algorithm and in each iteration, we ensure that the trajectory starts from the correct point on the invariant manifold, assuming the dynamics are given locally by (\ref{eq:tangent}), as described in algorithm \ref{algo:torusstability}.

\begin{algorithm}
\label{algo:torusstability}
\caption{Computation of the leading Lyapunov exponents of an invariant torus. Compare with algorithm \ref{algo:Lyapunov_exponents}.}
\SetKwInOut{Input}{Input}
\SetKwInOut{Output}{Output}
\Input{Parameters $L$, $\beta$ and $\epsilon$\\
Invariant torus $u(x,\rho,\sigma), R, S$. \\
Number of Lyapunov exponents $p$\\
Reorthonormalization time $\tau$\\
Number of reorthonormalizations $n$}
\Output{$p$ leading Lyapunov exponents $\mathbf{X}=[\chi_1\;\dots\;\chi_p]^\top$}
\vspace{2mm}\hrule\vspace{1mm}
$\tilde{\mathbf{Q}} \gets \left[\mathbf{I}_p \;|\; \mathbf{0}_{p\times(N-p)}\right]^\top$\;
QR decompose $\tilde{\mathbf{Q}}=\mathbf{Q}\mathbf{R}$\;
$\mathbf{X}\gets\mathbf{0}_{p\times1}$\;
\For{$j=1$ to $n$}{
    $u_0 \gets u(x,jR\tau ,jS\tau)$\;
    $\mathbf{W} \gets \mathbb{J}_{u_0}^\tau\mathbf{Q}$ \;
    QR decompose $\mathbf{W}=\mathbf{Q}\mathbf{R}$\;
    $\mathbf{X}\gets\mathbf{X}+\ln\left(\mathrm{diag}(\mathbf{R})\right)/\tau$\;
}
\Return $\mathbf{X}/n$
\end{algorithm}

All the UITs found had either two unstable eigenvalues (distinct values or repeated values, the latter suggesting a complex conjugate pair), or only one unstable eigenvalue. 
Any invariant solution embedded within the chaotic attractor can have at most two unstable eigenvalues, since this is the number of positive Lyapunov exponents for the attractor itself, as discussed in section \ref{sec:equations}.
Any UIT should have, by definition, two zero eigenvalues, and in each case we numerically found two Lyapunov exponents with magnitude very close to zero. The same method applied to UPOs gives just one (approximately) zero eigenvalue, and one or two positive ones. The full results are listed in table \ref{tab:weights}.

\section{Statistical predictions}
\label{sec:stats}

Let $\Gamma(u)$ be some real scalar measurable quantity for which we wish to know the long-time average for the system (\ref{eq:fgKSE}), for example the energy $\int_0^L \frac{1}{2}u^2 \mathrm{d}x$. For a trajectory confined to the $i$th UIT $u(x,\rho,\sigma)$ we may write, via Fourier transforms,
\begin{equation*}
    \Gamma(u(\cdot,\rho,\sigma)) = \sum_{k_1=-\infty}^\infty \sum_{k_2=-\infty}^\infty \Gamma^{(k_1,k_2)} e^{i(k_1 \rho+k_2 \sigma)} + \mathrm{c.c.}
\end{equation*}
and thus for a trajectory starting at $u(x,0,0)$, so that $\rho=Rt$ and $\sigma=St$, the long time average of $\Gamma(u)$ is given by
\begin{align*}
    \Gamma_i&\equiv\lim_{t\to\infty}\frac{1}{t}\int_0^t \Gamma(u) \mathrm{d}t \\
&= \lim_{t\to\infty}\frac{1}{t}\int_0^t \sum_{k_1=-\infty}^\infty \sum_{k_2=-\infty}^\infty \Gamma^{(k_1,k_2)} e^{i(k_1 R+k_2 S)t} \mathrm{d}t\\ &\qquad+ \mathrm{c.c.} \\
&= \Gamma^{(0,0)}\\
&= \frac{1}{4\pi^2}\int_0^{2\pi}\mathrm{d}\rho\int_0^{2\pi}\mathrm{d}\sigma \Gamma(u(x,\rho,\sigma)),
\end{align*}
so long as the torus is quasiperiodic, i.e. $R$ and $S$ are incommensurate, so that $k_1 R+k_2 S=0$ only if $k_1=k_2=0$.
In other words, the average value of a quantity $\Gamma(u)$ over a trajectory on a quasiperiodic invariant torus is the average $\Gamma_i$ over the torus surface itself, calculated in the obvious way.

We then wish to calculate weights $w_i$ so that the average value for the full chaotic attractor can be approximated as
\begin{equation}
    \hat{\Gamma} = \frac{\sum_i w_i \Gamma_i}{\sum_i w_i},
\end{equation}
where crucially the $w_i$ are independent of the particular quantity of interest $\Gamma$.

In certain cases, weights for sums of UPOs can be rigorously derived \citep{lan2010cycle}. When UITs are considered, it is not clear whether it is possible to derive an expression for weights analytically, though in our case it would be possible to use the weights for the corresponding RPOs in the unforced $\epsilon=0$ system.
 Even when a large number of periodic orbits are available, ad-hoc choices of weights have been found to give comparably good or even better results than these derivable weights \citep{chandler2013invariant}. The easiest choice of weights to adapt to UITs is that of \citet{zoldi1998spatially}, which is to assign the $i$th invariant solution a weight
\begin{equation}
\label{eq:weight}
    w_i=\frac{1}{\sum_{k:\chi^{(i)}_k>0}\chi^{(i)}_k},
\end{equation}
based on its positive Lyapunov exponents.
In the full periodic orbit theory, equilibria are excluded and only periodic orbits considered. An intuitive interpretation of this is that the UPOs have a greater `presence' in state space.
Extending this, we exclude the UPOs we have found from our calculations, and consider only UITs. Since all our UITs have either one or two unstable directions, (\ref{eq:weight}) reduces to
\begin{equation*}
w_i = \frac{1}{\chi^{(i)}_1+\chi^{(i)}_2},
\end{equation*}
where $\chi^{(i)}_2$ may be either zero or positive.

Tables \ref{tab:weights} and \ref{tab:weights2} list the UITs that were successfully converged at $\epsilon=0$ and continued to $\epsilon=0.001$ and $\epsilon=0.002$ respectively. The calculated Lyapunov exponents $\chi_1$ and $\chi_2$ for each of them are used to compute weights, and these weights are used to give predictions for the chaotic attractor for the energy production/dissipation, the Lyapunov exponents, and the first three Fourier coefficients. The values computed directly from a very long chaotic trajectory are also given, and the agreement is generally good.

\begin{table*}
    \centering
    \caption{For each of the twelve tori at $L=22$, $\epsilon=0.001$ and $\beta=0.01$, we list the energy production, the first two Lyapunov exponents and the absolute value of the first three Fourier coefficients. The Lyapunov exponents are used to calculate a weight for each torus, and these are then used to compute predictions for the quantities for the chaotic attractor. The values measured from a long time series on the chaotic attractor are also given. }
    \begin{tabular}{l | c c c c c c | c  }
        \toprule
Torus $T_i$ & $P=D$ & $\chi_1$ & $\chi_2$ & $|u^{(1)}|$ & $|u^{(2)}|$ & $|u^{(3)}|$ & $w_i/\sum_j w_j$ \\
        \midrule
        $T_1$&20.995&0.053&0       &4.350&13.428&10.099 &0.118\\
        $T_2$&19.218&0.032&0.032   &4.255&14.771&8.909  &0.098\\
        $T_3$&21.749&0.096&0       &4.694&12.492&11.109 &0.065\\
        $T_4$&20.235&0.062&0       &4.455&13.319&9.551  &0.102\\
        $T_5$&20.319&0.044&0       &4.353&13.813&9.626  &0.143\\
        $T_6$&21.401&0.032&0.029   &4.082&12.928&10.046 &0.104\\
        $T_7$&17.443&0.112&0       &5.196&11.394&8.416  &0.056\\
        $T_8$&24.216&0.056&0.050   &3.825&11.176&11.463 &0.059\\
        $T_9$&19.695&0.076&0       &4.556&12.727&9.388  &0.083\\
        $T_{10}$&23.632&0.325&0    &6.844&7.178&14.553  &0.019\\
        $T_{11}$&15.737&0.083&0.029&4.740&11.854&7.244  &0.056\\
        $T_{12}$&20.047&0.062&0.002&4.469&13.376&9.362  &0.098\\
        \midrule
        Prediction & 20.287 & 0.064 & 0.011 & 4.461 & 12.973 & 9.702 \\
        Chaotic attractor & 19.973 & 0.047 & 0.003 & 4.269 & 12.852 & 9.263 \\
        \bottomrule
    \end{tabular}
    \label{tab:weights}
\end{table*}

\begin{table*}
    \centering
    \caption{As for table \ref{tab:weights} but at $\epsilon=0.002$. Only 8 of the 12 UITs found at $\epsilon=0.001$ were successfully continued here.}
    \begin{tabular}{l | c c c c c c | c  }
        \toprule
Torus $T_i$ & $P=D$ & $\chi_1$ & $\chi_2$ & $|u^{(1)}|$ & $|u^{(2)}|$ & $|u^{(3)}|$ & $w_i/\sum_j w_j$ \\
        \midrule
        $T_2$&19.229&0.032&0.032   & 4.258&14.770&8.920 &0.151\\
        $T_5$&20.536&0.040&0       & 4.377&13.715&9.800 &0.239\\
        $T_6$&21.453&0.033&0.028   & 4.062&12.884&10.064&0.157\\
        $T_7$&17.661&0.114&0       & 5.139&11.325&8.518 &0.084\\
        $T_8$&24.382&0.054&0.050   & 3.814&11.118&11.567&0.092\\
        $T_{10}$&23.622&0.325&0    & 6.846&7.177&14.549 &0.030\\
        $T_{11}$&15.784&0.080&0.035& 4.773&11.862&7.296 &0.083\\
        $T_{12}$&20.121&0.056&0.002& 4.484&13.336&9.423 &0.164\\
        \midrule
        Prediction        & 20.222 & 0.060 & 0.017 & 4.445 & 12.893 & 9.633 \\
        Chaotic attractor & 20.030 & 0.045 & 0.004 & 4.231 & 12.948 & 9.252 \\
        \bottomrule
    \end{tabular}
    \label{tab:weights2}
\end{table*}

\section{Discussion}
\label{sec:conclusion}

We have demonstrated that, for this contrived system at carefully chosen parameter values, UITs are common and readily found. In addition to visually giving a good representation of the strange attractor, as shown in figure \ref{fig:alltori}, we are able to accurately predict statistics of the dynamics using a weighted average of the tori, with weights derived from the stabilities of the UITs.

The predictions of energy production and dissipation given in table \ref{tab:weights} are particularly good, despite the fact that the values for the individual UITs vary drastically.
The predicted values of the Lyapunov exponents of the chaotic attractor are significantly poorer, which is to be expected as these are notoriously difficult to calculate and sensitive to which parts of the attractor are accounted for.

It is self-evident that if UITs exist within an attractor, chaotic trajectories must pass very close to them and therefore, given sufficiently many UITs, the properties of the UITs can predict properties of the attractor. This system was chosen to give a method to find large numbers of UITs. Undoubtedly, UPOs still exist in this system as phase-lockings on the tori, but in general these are of long period.
What remains an open question is whether UITs exist in large numbers in realistic systems for which it has been difficult to detect UPOs, such as fluid turbulence. If they do, they could capture physical processes that are currently poorly understood. However, their detection, as well as the computational power needed to converge them, is a significant barrier to extend these ideas to such systems.

\section*{Acknowledgements}
This work was supported by the European Research Council (ERC) under the European Union's Horizon 2020 research and innovation programme (grant no. 865677).

\section*{Author Declarations}
The authors have no conflicts to disclose.

\section*{Data Availability}
The data that supports the findings of this study are available from the authors on request.

\section*{References}
\bibliography{references}

\begin{thebibliography}{29}%
\makeatletter
\providecommand \@ifxundefined [1]{%
 \@ifx{#1\undefined}
}%
\providecommand \@ifnum [1]{%
 \ifnum #1\expandafter \@firstoftwo
 \else \expandafter \@secondoftwo
 \fi
}%
\providecommand \@ifx [1]{%
 \ifx #1\expandafter \@firstoftwo
 \else \expandafter \@secondoftwo
 \fi
}%
\providecommand \natexlab [1]{#1}%
\providecommand \enquote  [1]{``#1''}%
\providecommand \bibnamefont  [1]{#1}%
\providecommand \bibfnamefont [1]{#1}%
\providecommand \citenamefont [1]{#1}%
\providecommand \href@noop [0]{\@secondoftwo}%
\providecommand \href [0]{\begingroup \@sanitize@url \@href}%
\providecommand \@href[1]{\@@startlink{#1}\@@href}%
\providecommand \@@href[1]{\endgroup#1\@@endlink}%
\providecommand \@sanitize@url [0]{\catcode `\\12\catcode `\$12\catcode
  `\&12\catcode `\#12\catcode `\^12\catcode `\_12\catcode `\%12\relax}%
\providecommand \@@startlink[1]{}%
\providecommand \@@endlink[0]{}%
\providecommand \url  [0]{\begingroup\@sanitize@url \@url }%
\providecommand \@url [1]{\endgroup\@href {#1}{\urlprefix }}%
\providecommand \urlprefix  [0]{URL }%
\providecommand \Eprint [0]{\href }%
\providecommand \doibase [0]{https://doi.org/}%
\providecommand \selectlanguage [0]{\@gobble}%
\providecommand \bibinfo  [0]{\@secondoftwo}%
\providecommand \bibfield  [0]{\@secondoftwo}%
\providecommand \translation [1]{[#1]}%
\providecommand \BibitemOpen [0]{}%
\providecommand \bibitemStop [0]{}%
\providecommand \bibitemNoStop [0]{.\EOS\space}%
\providecommand \EOS [0]{\spacefactor3000\relax}%
\providecommand \BibitemShut  [1]{\csname bibitem#1\endcsname}%
\let\auto@bib@innerbib\@empty
\bibitem [{\citenamefont {Pugh}(1967)}]{pugh1967improved}%
  \BibitemOpen
  \bibfield  {author} {\bibinfo {author} {\bibfnamefont {C.~C.}\ \bibnamefont
  {Pugh}},\ }\bibfield  {title} {\enquote {\bibinfo {title} {An improved
  closing lemma and a general density theorem},}\ }\href@noop {} {\bibfield
  {journal} {\bibinfo  {journal} {American Journal of Mathematics}\ }\textbf
  {\bibinfo {volume} {89}},\ \bibinfo {pages} {1010--1021} (\bibinfo {year}
  {1967})}\BibitemShut {NoStop}%
\bibitem [{\citenamefont {Bogomolny}(1992)}]{bogomolny1992dynamical}%
  \BibitemOpen
  \bibfield  {author} {\bibinfo {author} {\bibfnamefont {E.}~\bibnamefont
  {Bogomolny}},\ }\bibfield  {title} {\enquote {\bibinfo {title} {On dynamical
  zeta function},}\ }\href@noop {} {\bibfield  {journal} {\bibinfo  {journal}
  {Chaos: An Interdisciplinary Journal of Nonlinear Science}\ }\textbf
  {\bibinfo {volume} {2}},\ \bibinfo {pages} {5--13} (\bibinfo {year}
  {1992})}\BibitemShut {NoStop}%
\bibitem [{\citenamefont {Cvitanovi{\'c}}(1988)}]{cvitanovic1988invariant}%
  \BibitemOpen
  \bibfield  {author} {\bibinfo {author} {\bibfnamefont {P.}~\bibnamefont
  {Cvitanovi{\'c}}},\ }\bibfield  {title} {\enquote {\bibinfo {title}
  {Invariant measurement of strange sets in terms of cycles},}\ }\href@noop {}
  {\bibfield  {journal} {\bibinfo  {journal} {Physical Review Letters}\
  }\textbf {\bibinfo {volume} {61}},\ \bibinfo {pages} {2729} (\bibinfo {year}
  {1988})}\BibitemShut {NoStop}%
\bibitem [{\citenamefont {Eckhardt}\ and\ \citenamefont
  {Ott}(1994)}]{eckhardt1994periodic}%
  \BibitemOpen
  \bibfield  {author} {\bibinfo {author} {\bibfnamefont {B.}~\bibnamefont
  {Eckhardt}}\ and\ \bibinfo {author} {\bibfnamefont {G.}~\bibnamefont {Ott}},\
  }\bibfield  {title} {\enquote {\bibinfo {title} {Periodic orbit analysis of
  the {Lorenz} attractor},}\ }\href@noop {} {\bibfield  {journal} {\bibinfo
  {journal} {Zeitschrift f{\"u}r Physik B Condensed Matter}\ }\textbf {\bibinfo
  {volume} {93}},\ \bibinfo {pages} {259--266} (\bibinfo {year}
  {1994})}\BibitemShut {NoStop}%
\bibitem [{\citenamefont {Chandler}\ and\ \citenamefont
  {Kerswell}(2013)}]{chandler2013invariant}%
  \BibitemOpen
  \bibfield  {author} {\bibinfo {author} {\bibfnamefont {G.~J.}\ \bibnamefont
  {Chandler}}\ and\ \bibinfo {author} {\bibfnamefont {R.~R.}\ \bibnamefont
  {Kerswell}},\ }\bibfield  {title} {\enquote {\bibinfo {title} {Invariant
  recurrent solutions embedded in a turbulent two-dimensional kolmogorov
  flow},}\ }\href@noop {} {\bibfield  {journal} {\bibinfo  {journal} {Journal
  of Fluid Mechanics}\ }\textbf {\bibinfo {volume} {722}},\ \bibinfo {pages}
  {554--595} (\bibinfo {year} {2013})}\BibitemShut {NoStop}%
\bibitem [{\citenamefont {Anosov}\ and\ \citenamefont
  {Zhuzhoma}(2012)}]{anosov2012closing}%
  \BibitemOpen
  \bibfield  {author} {\bibinfo {author} {\bibfnamefont {D.}~\bibnamefont
  {Anosov}}\ and\ \bibinfo {author} {\bibfnamefont {E.}~\bibnamefont
  {Zhuzhoma}},\ }\bibfield  {title} {\enquote {\bibinfo {title} {Closing
  lemmas},}\ }\href@noop {} {\bibfield  {journal} {\bibinfo  {journal}
  {Differential equations}\ }\textbf {\bibinfo {volume} {48}},\ \bibinfo
  {pages} {1653--1699} (\bibinfo {year} {2012})}\BibitemShut {NoStop}%
\bibitem [{\citenamefont {Budanur}\ \emph {et~al.}(2017)\citenamefont
  {Budanur}, \citenamefont {Short}, \citenamefont {Farazmand}, \citenamefont
  {Willis},\ and\ \citenamefont {Cvitanovi{\'c}}}]{budanur2017relative}%
  \BibitemOpen
  \bibfield  {author} {\bibinfo {author} {\bibfnamefont {N.~B.}\ \bibnamefont
  {Budanur}}, \bibinfo {author} {\bibfnamefont {K.~Y.}\ \bibnamefont {Short}},
  \bibinfo {author} {\bibfnamefont {M.}~\bibnamefont {Farazmand}}, \bibinfo
  {author} {\bibfnamefont {A.~P.}\ \bibnamefont {Willis}},\ and\ \bibinfo
  {author} {\bibfnamefont {P.}~\bibnamefont {Cvitanovi{\'c}}},\ }\bibfield
  {title} {\enquote {\bibinfo {title} {Relative periodic orbits form the
  backbone of turbulent pipe flow},}\ }\href@noop {} {\bibfield  {journal}
  {\bibinfo  {journal} {Journal of Fluid Mechanics}\ }\textbf {\bibinfo
  {volume} {833}},\ \bibinfo {pages} {274--301} (\bibinfo {year}
  {2017})}\BibitemShut {NoStop}%
\bibitem [{\citenamefont {Parker}\ and\ \citenamefont
  {Schneider}(2022{\natexlab{a}})}]{parker2022invariant}%
  \BibitemOpen
  \bibfield  {author} {\bibinfo {author} {\bibfnamefont {J.~P.}\ \bibnamefont
  {Parker}}\ and\ \bibinfo {author} {\bibfnamefont {T.~M.}\ \bibnamefont
  {Schneider}},\ }\bibfield  {title} {\enquote {\bibinfo {title} {Invariant
  tori in dissipative hyperchaos},}\ }\href {https://doi.org/10.1063/5.0119642}
  {\bibfield  {journal} {\bibinfo  {journal} {Chaos}\ }\textbf {\bibinfo
  {volume} {32}},\ \bibinfo {pages} {113102} (\bibinfo {year}
  {2022}{\natexlab{a}})},\ \Eprint
  {https://arxiv.org/abs/https://doi.org/10.1063/5.0119642}
  {https://doi.org/10.1063/5.0119642} \BibitemShut {NoStop}%
\bibitem [{\citenamefont {Doohan}\ \emph {et~al.}(2022)\citenamefont {Doohan},
  \citenamefont {Bengana}, \citenamefont {Yang}, \citenamefont {Willis},\ and\
  \citenamefont {Hwang}}]{doohan2022state}%
  \BibitemOpen
  \bibfield  {author} {\bibinfo {author} {\bibfnamefont {P.}~\bibnamefont
  {Doohan}}, \bibinfo {author} {\bibfnamefont {Y.}~\bibnamefont {Bengana}},
  \bibinfo {author} {\bibfnamefont {Q.}~\bibnamefont {Yang}}, \bibinfo {author}
  {\bibfnamefont {A.~P.}\ \bibnamefont {Willis}},\ and\ \bibinfo {author}
  {\bibfnamefont {Y.}~\bibnamefont {Hwang}},\ }\bibfield  {title} {\enquote
  {\bibinfo {title} {The state space and travelling-wave solutions in two-scale
  wall-bounded turbulence},}\ }\href@noop {} {\bibfield  {journal} {\bibinfo
  {journal} {Journal of Fluid Mechanics}\ }\textbf {\bibinfo {volume} {947}},\
  \bibinfo {pages} {A41} (\bibinfo {year} {2022})}\BibitemShut {NoStop}%
\bibitem [{\citenamefont {Kudryashov}(1990)}]{kudryashov1990exact}%
  \BibitemOpen
  \bibfield  {author} {\bibinfo {author} {\bibfnamefont {N.~A.}\ \bibnamefont
  {Kudryashov}},\ }\bibfield  {title} {\enquote {\bibinfo {title} {Exact
  solutions of the generalized kuramoto-sivashinsky equation},}\ }\href@noop {}
  {\bibfield  {journal} {\bibinfo  {journal} {Physics Letters A}\ }\textbf
  {\bibinfo {volume} {147}},\ \bibinfo {pages} {287--291} (\bibinfo {year}
  {1990})}\BibitemShut {NoStop}%
\bibitem [{\citenamefont {Khater}\ and\ \citenamefont
  {Temsah}(2008)}]{khater2008numerical}%
  \BibitemOpen
  \bibfield  {author} {\bibinfo {author} {\bibfnamefont {A.}~\bibnamefont
  {Khater}}\ and\ \bibinfo {author} {\bibfnamefont {R.}~\bibnamefont
  {Temsah}},\ }\bibfield  {title} {\enquote {\bibinfo {title} {Numerical
  solutions of the generalized kuramoto--sivashinsky equation by chebyshev
  spectral collocation methods},}\ }\href@noop {} {\bibfield  {journal}
  {\bibinfo  {journal} {Computers \& Mathematics with Applications}\ }\textbf
  {\bibinfo {volume} {56}},\ \bibinfo {pages} {1465--1472} (\bibinfo {year}
  {2008})}\BibitemShut {NoStop}%
\bibitem [{\citenamefont {Lai}\ and\ \citenamefont
  {Ma}(2009)}]{lai2009lattice}%
  \BibitemOpen
  \bibfield  {author} {\bibinfo {author} {\bibfnamefont {H.}~\bibnamefont
  {Lai}}\ and\ \bibinfo {author} {\bibfnamefont {C.}~\bibnamefont {Ma}},\
  }\bibfield  {title} {\enquote {\bibinfo {title} {{Lattice Boltzmann method
  for the generalized Kuramoto--Sivashinsky equation}},}\ }\href@noop {}
  {\bibfield  {journal} {\bibinfo  {journal} {Physica A: Statistical Mechanics
  and its applications}\ }\textbf {\bibinfo {volume} {388}},\ \bibinfo {pages}
  {1405--1412} (\bibinfo {year} {2009})}\BibitemShut {NoStop}%
\bibitem [{\citenamefont {Lakestani}\ and\ \citenamefont
  {Dehghan}(2012)}]{lakestani2012numerical}%
  \BibitemOpen
  \bibfield  {author} {\bibinfo {author} {\bibfnamefont {M.}~\bibnamefont
  {Lakestani}}\ and\ \bibinfo {author} {\bibfnamefont {M.}~\bibnamefont
  {Dehghan}},\ }\bibfield  {title} {\enquote {\bibinfo {title} {Numerical
  solutions of the generalized kuramoto--sivashinsky equation using b-spline
  functions},}\ }\href@noop {} {\bibfield  {journal} {\bibinfo  {journal}
  {Applied Mathematical Modelling}\ }\textbf {\bibinfo {volume} {36}},\
  \bibinfo {pages} {605--617} (\bibinfo {year} {2012})}\BibitemShut {NoStop}%
\bibitem [{\citenamefont {LaQuey}\ \emph {et~al.}(1975)\citenamefont {LaQuey},
  \citenamefont {Mahajan}, \citenamefont {Rutherford},\ and\ \citenamefont
  {Tang}}]{laquey1975nonlinear}%
  \BibitemOpen
  \bibfield  {author} {\bibinfo {author} {\bibfnamefont {R.~E.}\ \bibnamefont
  {LaQuey}}, \bibinfo {author} {\bibfnamefont {S.}~\bibnamefont {Mahajan}},
  \bibinfo {author} {\bibfnamefont {P.}~\bibnamefont {Rutherford}},\ and\
  \bibinfo {author} {\bibfnamefont {W.}~\bibnamefont {Tang}},\ }\bibfield
  {title} {\enquote {\bibinfo {title} {Nonlinear saturation of the trapped-ion
  mode},}\ }\href@noop {} {\bibfield  {journal} {\bibinfo  {journal} {Physical
  Review Letters}\ }\textbf {\bibinfo {volume} {34}},\ \bibinfo {pages} {391}
  (\bibinfo {year} {1975})}\BibitemShut {NoStop}%
\bibitem [{\citenamefont {Sivashinsky}(1977)}]{sivashinsky1977nonlinear}%
  \BibitemOpen
  \bibfield  {author} {\bibinfo {author} {\bibfnamefont {G.~I.}\ \bibnamefont
  {Sivashinsky}},\ }\bibfield  {title} {\enquote {\bibinfo {title} {Nonlinear
  analysis of hydrodynamic instability in laminar flames—i. derivation of
  basic equations},}\ }\href@noop {} {\bibfield  {journal} {\bibinfo  {journal}
  {Acta astronautica}\ }\textbf {\bibinfo {volume} {4}},\ \bibinfo {pages}
  {1177--1206} (\bibinfo {year} {1977})}\BibitemShut {NoStop}%
\bibitem [{\citenamefont {Chang}(1994)}]{chang1994wave}%
  \BibitemOpen
  \bibfield  {author} {\bibinfo {author} {\bibfnamefont {H.}~\bibnamefont
  {Chang}},\ }\bibfield  {title} {\enquote {\bibinfo {title} {Wave evolution on
  a falling film},}\ }\href@noop {} {\bibfield  {journal} {\bibinfo  {journal}
  {Annual review of fluid mechanics}\ }\textbf {\bibinfo {volume} {26}},\
  \bibinfo {pages} {103--136} (\bibinfo {year} {1994})}\BibitemShut {NoStop}%
\bibitem [{\citenamefont {Kassam}\ and\ \citenamefont
  {Trefethen}(2005)}]{kassam2005fourth}%
  \BibitemOpen
  \bibfield  {author} {\bibinfo {author} {\bibfnamefont {A.-K.}\ \bibnamefont
  {Kassam}}\ and\ \bibinfo {author} {\bibfnamefont {L.~N.}\ \bibnamefont
  {Trefethen}},\ }\bibfield  {title} {\enquote {\bibinfo {title} {Fourth-order
  time-stepping for stiff pdes},}\ }\href@noop {} {\bibfield  {journal}
  {\bibinfo  {journal} {SIAM Journal on Scientific Computing}\ }\textbf
  {\bibinfo {volume} {26}},\ \bibinfo {pages} {1214--1233} (\bibinfo {year}
  {2005})}\BibitemShut {NoStop}%
\bibitem [{\citenamefont {Innes}(2018)}]{innes2018don}%
  \BibitemOpen
  \bibfield  {author} {\bibinfo {author} {\bibfnamefont {M.}~\bibnamefont
  {Innes}},\ }\bibfield  {title} {\enquote {\bibinfo {title} {Don't unroll
  adjoint: Differentiating ssa-form programs},}\ }\href@noop {} {\bibfield
  {journal} {\bibinfo  {journal} {arXiv preprint arXiv:1810.07951}\ } (\bibinfo
  {year} {2018})}\BibitemShut {NoStop}%
\bibitem [{\citenamefont {Benettin}\ \emph {et~al.}(1980)\citenamefont
  {Benettin}, \citenamefont {Galgani}, \citenamefont {Giorgilli},\ and\
  \citenamefont {Strelcyn}}]{Benettin1980}%
  \BibitemOpen
  \bibfield  {author} {\bibinfo {author} {\bibfnamefont {G.}~\bibnamefont
  {Benettin}}, \bibinfo {author} {\bibfnamefont {L.}~\bibnamefont {Galgani}},
  \bibinfo {author} {\bibfnamefont {A.}~\bibnamefont {Giorgilli}},\ and\
  \bibinfo {author} {\bibfnamefont {J.-M.}\ \bibnamefont {Strelcyn}},\
  }\bibfield  {title} {\enquote {\bibinfo {title} {{Lyapunov Characteristic
  Exponents for smooth dynamical systems and for hamiltonian systems; A method
  for computing all of them. Part 2: Numerical application}},}\ }\href
  {https://doi.org/10.1007/BF02128237} {\bibfield  {journal} {\bibinfo
  {journal} {Meccanica}\ }\textbf {\bibinfo {volume} {15}},\ \bibinfo {pages}
  {21--30} (\bibinfo {year} {1980})}\BibitemShut {NoStop}%
\bibitem [{\citenamefont {Eckmann}\ and\ \citenamefont
  {Ruelle}(1985)}]{Eckmann1985}%
  \BibitemOpen
  \bibfield  {author} {\bibinfo {author} {\bibfnamefont {J.~P.}\ \bibnamefont
  {Eckmann}}\ and\ \bibinfo {author} {\bibfnamefont {D.}~\bibnamefont
  {Ruelle}},\ }\bibfield  {title} {\enquote {\bibinfo {title} {Ergodic theory
  of chaos and strange attractors},}\ }\href
  {https://doi.org/10.1103/RevModPhys.57.617} {\bibfield  {journal} {\bibinfo
  {journal} {Rev. Mod. Phys.}\ }\textbf {\bibinfo {volume} {57}},\ \bibinfo
  {pages} {617--656} (\bibinfo {year} {1985})}\BibitemShut {NoStop}%
\bibitem [{\citenamefont {Cvitanovi{\'c}}, \citenamefont {Davidchack},\ and\
  \citenamefont {Siminos}(2010)}]{cvitanovic2010state}%
  \BibitemOpen
  \bibfield  {author} {\bibinfo {author} {\bibfnamefont {P.}~\bibnamefont
  {Cvitanovi{\'c}}}, \bibinfo {author} {\bibfnamefont {R.~L.}\ \bibnamefont
  {Davidchack}},\ and\ \bibinfo {author} {\bibfnamefont {E.}~\bibnamefont
  {Siminos}},\ }\bibfield  {title} {\enquote {\bibinfo {title} {On the state
  space geometry of the kuramoto--sivashinsky flow in a periodic domain},}\
  }\href@noop {} {\bibfield  {journal} {\bibinfo  {journal} {SIAM Journal on
  Applied Dynamical Systems}\ }\textbf {\bibinfo {volume} {9}},\ \bibinfo
  {pages} {1--33} (\bibinfo {year} {2010})}\BibitemShut {NoStop}%
\bibitem [{\citenamefont {Fong}\ and\ \citenamefont
  {Saunders}(2011)}]{fong2011lsmr}%
  \BibitemOpen
  \bibfield  {author} {\bibinfo {author} {\bibfnamefont {D.~C.-L.}\
  \bibnamefont {Fong}}\ and\ \bibinfo {author} {\bibfnamefont {M.}~\bibnamefont
  {Saunders}},\ }\bibfield  {title} {\enquote {\bibinfo {title} {Lsmr: An
  iterative algorithm for sparse least-squares problems},}\ }\href@noop {}
  {\bibfield  {journal} {\bibinfo  {journal} {SIAM Journal on Scientific
  Computing}\ }\textbf {\bibinfo {volume} {33}},\ \bibinfo {pages} {2950--2971}
  (\bibinfo {year} {2011})}\BibitemShut {NoStop}%
\bibitem [{\citenamefont {Azimi}, \citenamefont {Ashtari},\ and\ \citenamefont
  {Schneider}(2022)}]{azimi2020adjoint}%
  \BibitemOpen
  \bibfield  {author} {\bibinfo {author} {\bibfnamefont {S.}~\bibnamefont
  {Azimi}}, \bibinfo {author} {\bibfnamefont {O.}~\bibnamefont {Ashtari}},\
  and\ \bibinfo {author} {\bibfnamefont {T.~M.}\ \bibnamefont {Schneider}},\
  }\bibfield  {title} {\enquote {\bibinfo {title} {Constructing periodic orbits
  of high-dimensional chaotic systems by an adjoint-based variational
  method},}\ }\href {https://doi.org/10.1103/PhysRevE.105.014217} {\bibfield
  {journal} {\bibinfo  {journal} {Phys. Rev. E}\ }\textbf {\bibinfo {volume}
  {105}},\ \bibinfo {pages} {014217} (\bibinfo {year} {2022})}\BibitemShut
  {NoStop}%
\bibitem [{\citenamefont {Parker}\ and\ \citenamefont
  {Schneider}(2022{\natexlab{b}})}]{parker2022variational}%
  \BibitemOpen
  \bibfield  {author} {\bibinfo {author} {\bibfnamefont {J.~P.}\ \bibnamefont
  {Parker}}\ and\ \bibinfo {author} {\bibfnamefont {T.~M.}\ \bibnamefont
  {Schneider}},\ }\bibfield  {title} {\enquote {\bibinfo {title} {Variational
  methods for finding periodic orbits in the incompressible navier--stokes
  equations},}\ }\href@noop {} {\bibfield  {journal} {\bibinfo  {journal}
  {Journal of Fluid Mechanics}\ }\textbf {\bibinfo {volume} {941}} (\bibinfo
  {year} {2022}{\natexlab{b}})}\BibitemShut {NoStop}%
\bibitem [{\citenamefont {Page}\ \emph {et~al.}(2022)\citenamefont {Page},
  \citenamefont {Norgaard}, \citenamefont {Brenner},\ and\ \citenamefont
  {Kerswell}}]{page2022recurrent}%
  \BibitemOpen
  \bibfield  {author} {\bibinfo {author} {\bibfnamefont {J.}~\bibnamefont
  {Page}}, \bibinfo {author} {\bibfnamefont {P.}~\bibnamefont {Norgaard}},
  \bibinfo {author} {\bibfnamefont {M.~P.}\ \bibnamefont {Brenner}},\ and\
  \bibinfo {author} {\bibfnamefont {R.~R.}\ \bibnamefont {Kerswell}},\
  }\bibfield  {title} {\enquote {\bibinfo {title} {Recurrent flow patterns as a
  basis for turbulence: predicting statistics from structures},}\ }\href@noop
  {} {\bibfield  {journal} {\bibinfo  {journal} {arXiv preprint
  arXiv:2212.11886}\ } (\bibinfo {year} {2022})}\BibitemShut {NoStop}%
\bibitem [{\citenamefont {Jorba}(2001)}]{jorba2001numerical}%
  \BibitemOpen
  \bibfield  {author} {\bibinfo {author} {\bibfnamefont {{\`A}.}~\bibnamefont
  {Jorba}},\ }\bibfield  {title} {\enquote {\bibinfo {title} {Numerical
  computation of the normal behaviour of invariant curves of n-dimensional
  maps},}\ }\href@noop {} {\bibfield  {journal} {\bibinfo  {journal}
  {Nonlinearity}\ }\textbf {\bibinfo {volume} {14}},\ \bibinfo {pages} {943}
  (\bibinfo {year} {2001})}\BibitemShut {NoStop}%
\bibitem [{\citenamefont {Wysham}\ and\ \citenamefont
  {Meiss}(2006)}]{wysham2006iterative}%
  \BibitemOpen
  \bibfield  {author} {\bibinfo {author} {\bibfnamefont {D.~B.}\ \bibnamefont
  {Wysham}}\ and\ \bibinfo {author} {\bibfnamefont {J.~D.}\ \bibnamefont
  {Meiss}},\ }\bibfield  {title} {\enquote {\bibinfo {title} {Iterative
  techniques for computing the linearized manifolds of quasiperiodic tori},}\
  }\href@noop {} {\bibfield  {journal} {\bibinfo  {journal} {Chaos: An
  Interdisciplinary Journal of Nonlinear Science}\ }\textbf {\bibinfo {volume}
  {16}},\ \bibinfo {pages} {023129} (\bibinfo {year} {2006})}\BibitemShut
  {NoStop}%
\bibitem [{\citenamefont {Lan}(2010)}]{lan2010cycle}%
  \BibitemOpen
  \bibfield  {author} {\bibinfo {author} {\bibfnamefont {Y.}~\bibnamefont
  {Lan}},\ }\bibfield  {title} {\enquote {\bibinfo {title} {Cycle expansions:
  From maps to turbulence},}\ }\href@noop {} {\bibfield  {journal} {\bibinfo
  {journal} {Communications in Nonlinear Science and Numerical Simulation}\
  }\textbf {\bibinfo {volume} {15}},\ \bibinfo {pages} {502--526} (\bibinfo
  {year} {2010})}\BibitemShut {NoStop}%
\bibitem [{\citenamefont {Zoldi}\ and\ \citenamefont
  {Greenside}(1998)}]{zoldi1998spatially}%
  \BibitemOpen
  \bibfield  {author} {\bibinfo {author} {\bibfnamefont {S.~M.}\ \bibnamefont
  {Zoldi}}\ and\ \bibinfo {author} {\bibfnamefont {H.~S.}\ \bibnamefont
  {Greenside}},\ }\bibfield  {title} {\enquote {\bibinfo {title} {Spatially
  localized unstable periodic orbits of a high-dimensional chaotic system},}\
  }\href@noop {} {\bibfield  {journal} {\bibinfo  {journal} {Physical Review
  E}\ }\textbf {\bibinfo {volume} {57}},\ \bibinfo {pages} {R2511} (\bibinfo
  {year} {1998})}\BibitemShut {NoStop}%
\end{thebibliography}%


%

\end{document}